\documentclass[conference]{IEEEtran}
\IEEEoverridecommandlockouts
\usepackage{cite}
\usepackage{amsmath,amssymb,amsfonts}
\usepackage{algorithmic}
\usepackage{graphicx}
\usepackage{textcomp}
\usepackage{url}
\usepackage{enumitem} 
\usepackage{hyperref}
\usepackage{xcolor}
\usepackage{float}
\usepackage{stfloats}
\usepackage[dvipsnames,table,svgnames]{xcolor}
\usepackage{xr-hyper}
\def\BibTeX{{\rm B\kern-.05em{\sc i\kern-.025em b}\kern-.08em
    T\kern-.1667em\lower.7ex\hbox{E}\kern-.125emX}}
\begin{document}

\title{Ranking Companion: A Visual Analytics Approach to Item-Based Ranking with Hybrid Item Selection}

\author{%
\IEEEauthorblockN{Aman Kumar\IEEEauthorrefmark{1},
Maximilian Tornow\IEEEauthorrefmark{1},
Michaela Benk\IEEEauthorrefmark{1},
Ibrahim Al-Hazwani\IEEEauthorrefmark{1}\IEEEauthorrefmark{2},
and J\"urgen Bernard\IEEEauthorrefmark{1}\IEEEauthorrefmark{2}}
\IEEEauthorblockA{%
\IEEEauthorrefmark{1}\textit{University of  Z\"urich}, Z\"urich, Switzerland \\
\IEEEauthorrefmark{2}\textit{Digital Society Initiative},  Z\"urich, Switzerland \\
Email: \{aman.kumar, maximilian.tornow, michaela.benk, ibrahim.alhazwani, juergen.bernard\}@uzh.ch}
}


\maketitle

\begin{figure*}[!t]
\centering
\includegraphics[width=\linewidth]{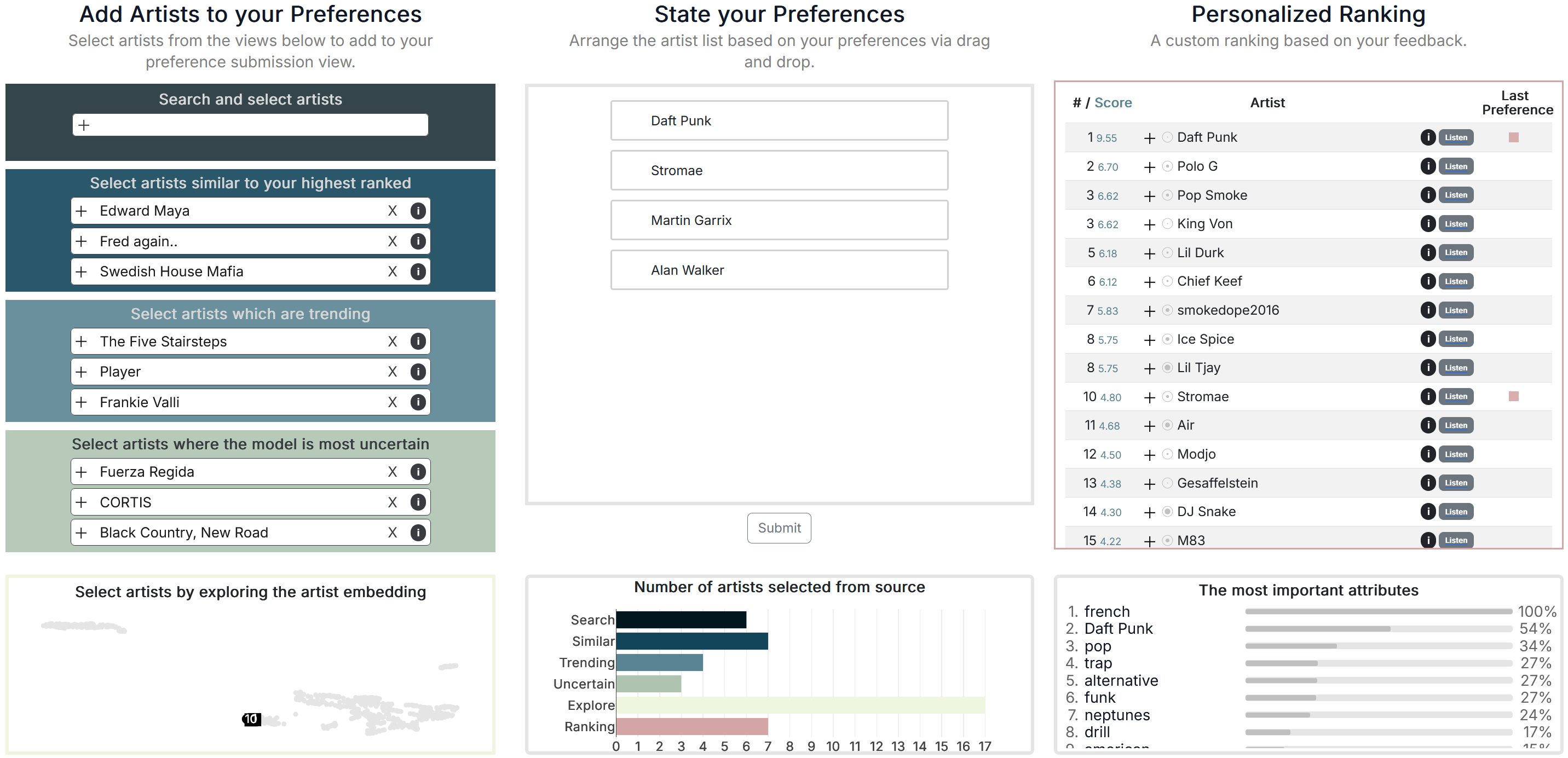}
\vspace{-2mm}
\caption{The Ranking Companion interface implements an iterative three-step ranking creation workflow: select (artist) items with different methods (left), externalize listwise preference feedback for ranking-model training (center), and inspect and validate the ranking results and their explanations (right).}
\label{fig:teaser}
\vspace{-5mm}
\end{figure*}

\begin{abstract}
Personalizing item ranking creation is a challenging task, especially when users lack knowledge of data attributes or the ability to express and formalize their attribute preferences. 
Item-based ranking creation is an approach allowing users to directly externalize preferences through known-item judgments rather than attribute-based scoring. 
However, a core challenge of item-based ranking is identifying and selecting representative candidate items for externalizing preferences.
Existing approaches rely on singular item-selection methods, limiting flexibility and user control. 
To address this challenge, we present \textit{Ranking Companion}, a visual analytics approach for item-based ranking that combines model-driven active learning with human-driven item-selection methods. 
By drawing from six complementary item-selection methods, users can externalize listwise preferences based on selected candidate items, while an iterative machine learning process with a ranking model calculates ranking results, presented to users alongside explanations for interpretation. 
We evaluated Ranking Companion in a formative user study $(N=10)$ in which participants used each item-selection method across three iterations, revealing tradeoffs in perceived ranking quality across accuracy, diversity, novelty, transparency, control, and satisfaction. 
Ranking Companion contributes a unified interactive item selection space and provides preliminary empirical guidance toward the hybrid use of multiple complementary item-selection methods in personalized item-based ranking creation.

\end{abstract}


\section{Introduction}

Item rankings are an important decision-making support mechanism in both professional and personal contexts. 
Their core strength lies in transforming large collections of alternatives into ordered lists by leveraging complex, multi-criteria information, enabling users to decide on high-priority, top-listed items. 
In professional domains, item rankings support decision-making such as hiring, supplier selection, university admissions, risk assessment, and investment analysis. 
In everyday settings, item rankings support choices in travel planning, media consumption, consumer products, and healthcare.
The creation of item rankings is inherently subjective, preference-driven, and personal, as individuals differ in their goals, circumstances, risk tolerance, and values.

Approaches to personalized ranking creation range from implicit to explicit feedback mechanisms, differing in how directly users can express their preferences. 
Their processes further range from automated black-box modeling to interactive systems that expose model behavior and invite user control.
\textit{Visual Analytics} (VA) approaches fall into the latter of these categories, with human-in-the-loop support for explicit item-based or attribute-based preference externalization, the ability to inspect and adjust attribute weights, and the analysis of interpretable model outcomes.
These tools provide user control, support for iterative and incremental learning cycles, and yield explainable outcomes~\cite{amershi_power_2014,dudley_review_2018,endert_human_2014}.
Our research investigates the efficacy of human-centered ranking creation, which depends largely on what users already know about the data, how well users can engage with the provided feedback mechanisms, and how accurate feedback interactions are transformed into mathematical ranking models.


\textit{Attribute-based ranking} approaches enable users to externalize preferences through attribute scoring functions \cite{schmid_taxonomy_2021} and explore the relationship between adjustable attribute weights and ranking results~\cite{pajer2016weightlifter}. 
Common examples include \emph{LineUp}~\cite{gratzl_lineup_2013}, \emph{RankASco}~\cite{schmid_rankasco_2022}, and PAVED~\cite{cibulskiMMK20}.
These tools excel when a) the attribute/feature space is semantically meaningful (as opposed to, e.g., deep-learned and/or embedding features), b) users are familiar with the attribute space, c) interfaces offer the types of scoring function users have in mind (which are then translated into a mathematical model), and d) users can articulate their preferences through meaningful interactions.
Prior work, however, highlights that attribute-based translations from users' minds to mathematical models are not trivial and may introduce burdens, especially for unknown attributes and for non-experts~\cite{barth_how_2023}.

\textit{Item-based ranking} reduces assumptions about users' familiarity with the attribute space by asking users to compare concrete item pairs or small subsets, which can be more reliable than specifying attribute preferences and weights~\cite{carterette_here_2008}.
A common approach is to leverage learn-to-rank models that generalize from these item-based preferences~\cite{liu_learning_2007}.
Research argues that ordered item lists offer richer training signals than pairwise feedback \cite{xia_listwise_2008} and that relative preference judgments are easier to obtain than absolute ratings \cite{carterette_here_2008,cibulskiMMK20}, motivating interfaces that capture relative listwise orderings directly.
Two VA approaches operationalize item feedback: Podium and RanKit. 
\emph{Podium} lets users drag items from a tabular ranking inspection view to express local ordering constraints, inferring attribute weights using a Ranking SVM model~\cite{wall_podium_2018}.
\emph{RanKit} avoids showing an initial ranking to reduce anchoring bias, letting users construct preferences via pairwise, listwise, or categorical comparisons, to incrementally learn a personalized ranker~\cite{kuhlman_preference-driven_2018}.

Despite this progress, a challenge of item-based ranking creation remains unaddressed: users or computational agents must decide which items to select next, as a basis for preference externalization.
These choices shape ranking quality by determining which preference evidence is available for model training~\cite{settles_active_2009,fu_survey_2013}. 
Podium exposes items through a single ranking view, while RanKit relies on manual browsing and repeated shuffling.
Both approaches offer only one fixed item-selection method, a limitation well-known in active learning research: relying on a single strategy risks missing informative regions of the data space~\cite{settles_active_2009}.
The goals of our research are threefold: (a) identify a diverse set of item-selection methods that effectively support ranking creation, (b) develop the first item-based ranking approach that combines these strategies within a unified workflow, and (c) evaluate their combined usefulness.
Our approach explicitly studies human-based and model-based item selection, as well as collaborative forms with different agency and control.

We contribute \textbf{Ranking Companion}, a VA approach for item-based preference externalization and ranking creation.
The approach is built around three coordinated interface components: item selection, listwise preference input, and ranking inspection, which together form a closed interactive machine learning loop.
Ranking Companion's main novelty is in its combination of multiple complementary item-selection methods, integrating \emph{Active Learning} (AL) alongside human-centric methods to improve ranking model performance, support the data space coverage, and reflect user preferences.
Figure~\ref{fig:teaser} shows the three coordinated views that implement a ranking creation workflow: item selection $\rightarrow$ externalize listwise preferences for modeling $\rightarrow$ ranking analysis.
We evaluate Ranking Companion in a music artist ranking case and report preliminary results of a user study, assessing how six item-selection methods affect ten users' perceived ranking quality.
Ranking Companion extends personalized item-based ranking creation with increased human control and agency beyond relying on any single item-selection method, and lays the groundwork for more studies of human-AI collaboration for item-based ranking creation.
Supplementary material and a video guide are available online.\footnote{\href{https://osf.io/x54t3/overview?view_only=ad39bb4b8e0446aeb4c960b4e1de581d}{Supplementary material and video guide.}}

\section{Ranking Companion Approach}
We start with a general data abstraction and an introduction to the running example on music artists and corresponding feature preprocessing.
At the core, we present the iterative ranking-creation workflow with its three tasks, defining the following three subsections on item-selection methods, listwise item feedback, and feedback interpretation. 

\subsection{Data Abstraction and Case}
\label{sec:approach:data}
We demonstrate the approach using \emph{music artist} \emph{items}, using a Last.fm dataset \cite{lastfm_ltd_about_nodate}.
Each item is characterized by user-generated attribute tags such as genres, countries, or contextual descriptors (e.g., ``seen live'').
The Last.fm API provides these attributes as relevance scores, normalized to $[0,100]$. 
We filter low-relevance attributes using a threshold of 10, following prior work on metadata quality \cite{bogdanov_how_2011}. 
Overall, we retrieved $6{,}000$ popular artists and associated attributes.
For the formative study, we used a subset of $1{,}000$ items to reduce training latency and maintain interactive response times.

\subsection{Iterative Workflow and Tasks}
\label{sec:approach:workflowTasks}

Ranking Companion's workflow emerges from four complementary methodological foundations. 
\textit{Interactive machine learning}~\cite{amershi_power_2014, dudley_review_2018} contributes the closed-loop structure of iterative feedback submission and model update, keeping the user in control of an incrementally evolving ranking model.
\textit{Interactive visual data labeling}~\cite{Seifert10User,Hoeferlin2012VIAL,bernard_vial_2018} provides the human-centric and iterative selection of items, user feedback, and feedback interpretation.
\textit{Active learning}~\cite{settles_active_2009, fu_survey_2013} motivates the diversity of item selection strategies, as no single strategy dominates across all criteria such as informativeness, coverage, and user comprehension.
\textit{Item-based ranking creation} motivates the core interactive feedback mechanism~\cite{wall_podium_2018,kuhlman_preference-driven_2018}: users express preferences over concrete items rather than abstract attributes. 
These four perspectives converge on three core tasks that structure the workflow.
\begin{enumerate}[leftmargin=*,itemsep=1pt, label=\textbf{T\arabic*:}]
\item \textbf{Select candidate items}, drawing from multiple methods that facilitate interactive-visual item selection. 
\item \textbf{Externalize listwise preferences} over a small set of items that the learning model can use for training. 
\item \textbf{Inspect and validate the ranking} and its explanations to decide whether the modeling aligns with user preferences.
\end{enumerate}
We describe how each task is realized: 
the item-selection methods (Section~\ref{sec:approach:itemSelection}), 
the listwise preference externalization mechanism (Section~\ref{sec:approach:itemFeedback}), 
and the feedback interpretation mechanism and learning-to-rank model (Section~\ref{sec:approach:model}).


\subsection{Item-Selection Methods} 
\label{sec:approach:itemSelection}

To diversify item selection, we opted for methods that cover complementary perspectives: human-centered selection, model-centered selection, data-centered exploration, dataset-intrinsic signals, and ranking-intrinsic refinement. 
Together, these criteria ensure that no single bias (whether from the user, the model, or the data) dominates the selection process.
We incorporate model-based AL methods for item selection, which reduces labeling effort by selecting informative items \cite{settles_active_2009,fu_survey_2013}.
In addition, we introduce interactive and visual human-based item-selection methods, adopted from conceptual process models~\cite{Seifert10User,Hoeferlin2012VIAL,bernard_vial_2018}, item selection taxonomies~\cite{bernard_taxonomy_2021}, empirical studies~\cite{bernard_towards_2018,tvcgLabeling2018,chegini2020}, and best practices~\cite{ritter_personalized_2018,tiis2021Sevastjanova}.
While prior research for item selection mainly targets classification and regression, we adopt these principles for ranking creation.
To meet all criteria, Ranking Companion offers six complementary item-selection methods: 

\textbf{ISM0 Baseline Search:} addresses a human-centered criterion: users retrieve items they already know by name~\cite{hearst_search_2009}, giving them agency and a familiar starting point for anchoring early preferences, mitigating bootstrap problems~\cite{settles_active_2009}.

\textbf{ISM1 Similarity:} bridges the human and model perspectives: items are suggested in proximity to already top-ranked items~\cite{bernard_taxonomy_2021}, exploiting known preference regions while remaining guided by the current model state.

\textbf{ISM2 Trend:} addresses the dataset-intrinsic criterion: rather than deriving suggestions from the user or the model, it suggests items based on a signal specific to the Last.fm dataset, introducing an external perspective independent of labeled history.
    We estimate the trend score of an item
    \(x\) as
    \begin{equation}
        T(x)=\frac{L_x}{P_x},
    \end{equation}
    where \(L_x\) denotes the current number of listeners and \(P_x\) denotes
    the total play count of item \(x\)~\cite{celma_music_2010}.

\textbf{ISM3 Uncertainty: } addresses a model-centered criterion inspired by active learning: items are suggested from regions where the current ranking model has received little preference evidence~\cite{settles_active_2009}, with the goal of improving the ranking model in each feedback round.

    Let \(p_a\) denote the frequency with
    which attribute \(a\) occurs among the items already labeled by the user,
    and let \(p_a=-1\) for previously unseen attributes. We then define
    \begin{equation}
        U(x)=\sum_{a \in A} w_{x,a}\, p_a,
        \label{eq:uncertainty}
    \end{equation}
    where \(A\) is the set of all attributes and \(w_{x,a}\) denotes the
    normalized relevance weight of attribute \(a\) for item \(x\) (with
    \(w_{x,a}=0\) if item \(x\) does not exhibit attribute \(a\)). Lower values of \(U(x)\) indicate that an item is characterized by attributes that are underrepresented in the current labeled history. We use this score as an interpretable proxy for model uncertainty: it does not estimate calibrated probabilistic uncertainty, but it suggests items from regions where the ranking model has received little preference evidence~\cite{settles_active_2009,fu_survey_2013}.

\textbf{ISM4 UMAP Projection Map: } addresses the data-centered criterion: by projecting the full item space into two dimensions~\cite{mcinnes_umap_2018}, it enables spatial browsing and pattern-driven discovery based on the structure of the item space itself~\cite{tvcgLabeling2018}, independent of the users' history and model state.

\textbf{ISM5 Ranking Panel:} addresses the ranking-intrinsic criterion: users add items directly from the current ranking result, enabling targeted correction and iterative refinement of specific placements based on visual model-outcome assessment~\cite{bernard_towards_2018}.
\subsection{Listwise Preference Externalization} 
\label{sec:approach:itemFeedback}

After candidate items have been selected through any of the six item-selection methods, Ranking Companion asks users to externalize preferences by ordering the selected items.
This separates the question of which items should be judged from the question of how users express preference judgments over those items.

\begin{figure}[t]
\centering
\includegraphics[width=\linewidth]{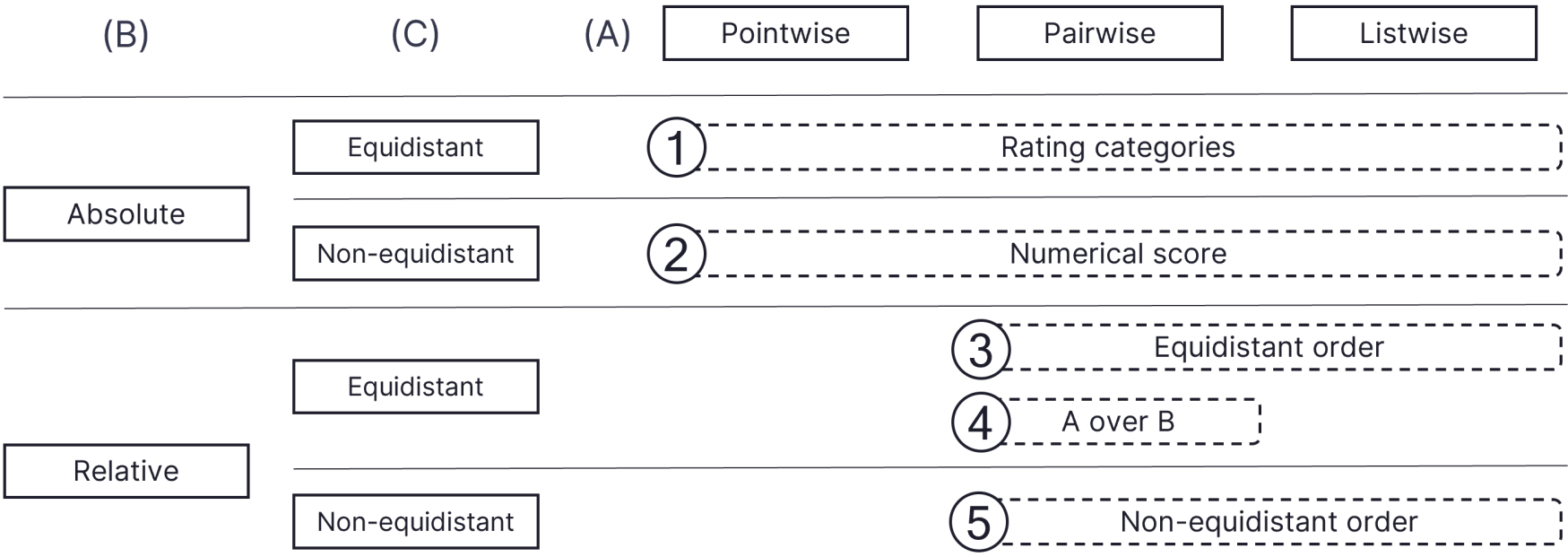}
\vspace{-2mm}
\caption{Preference-input mechanics. (A) number of items per statement, (B) type of statement, and (C) how distances between adjacent positions are interpreted.}
\label{fig:interaction-paradigms}
\vspace{-5mm}
\end{figure}

Item-based ranking interfaces differ in how they capture users' preferences. 
We characterize this preference-input design space along three independent dimensions (Figure~\ref{fig:interaction-paradigms}):
\begin{enumerate}[leftmargin=*,itemsep=1pt,topsep=2pt]
\item \textbf{Items per statement:} pointwise, pairwise, or listwise~\cite{liu_learning_2007}.
\item \textbf{Statement type:} absolute judgments versus relative comparisons~\cite{carterette_here_2008}.
\item \textbf{Distance encoding:} equal versus unequal spacing between adjacent positions~\cite{kuhlman_preference-driven_2018}.
\end{enumerate}

\begin{figure*}[!t]
    \centering
    \vspace{-3mm}
    \includegraphics[width=\textwidth]{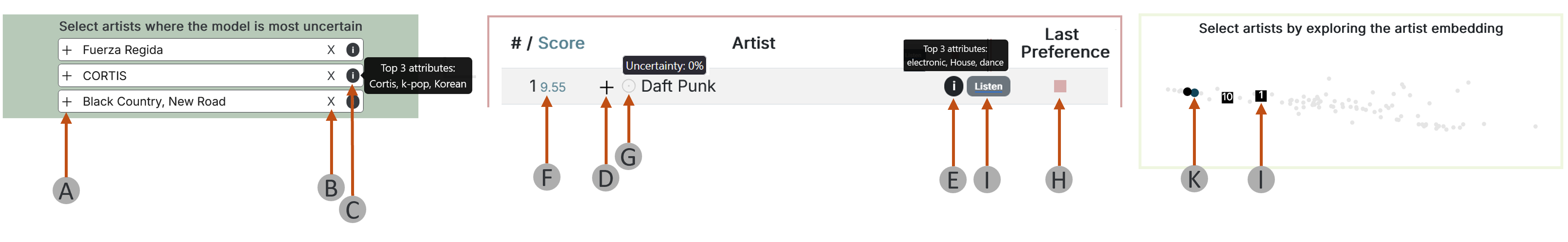}
    \vspace{-8mm}
    \caption{Designs for item-selection methods in Ranking Companion: row-based suggestion lists for ISM1--ISM3, projection-based selection for ISM4, and ranking-panel selection for ISM5. The interface provides eleven visual cues and controls, labeled A--I and K--L; see Section~\ref{sec:interface} for details.}
    \label{fig:itemSelectionViewTypes}
    \vspace{-2mm}
\end{figure*}


Ranking Companion's design combines three decisions: listwise, relative, and equidistant feedback.

Listwise feedback provides a richer supervision signal than isolated pairwise judgments because one ordered list contains multiple ordering relations and can be used directly as one training group for the learn-to-rank objective~\cite{xia_listwise_2008}. This makes it well-suited for iterative ranking creation, where each interaction should provide an informative training signal.

Relative feedback avoids asking users to assign absolute scores to items. Prior work shows that comparative preference judgments can be easier and more reliable than absolute ratings, especially when users judge concrete items rather than abstract attribute weights~\cite{carterette_here_2008,cibulskiMMK20}.

Equidistant spacing gives the ranking model a straightforward ordinal training signal: adjacent positions in the submitted list are interpreted as equally spaced relevance steps. This is a pragmatic and transparent default for a first prototype because it avoids additional interaction cost and keeps the mapping from drag-and-drop order to model input simple. We therefore treat equidistant spacing as a best-practice simplification for the current design, not as a claim that users' true preference strengths are always equally spaced.

\subsection{Feedback Interpretation and Learning-to-Rank Model}
\label{sec:approach:model}

Ranking Companion accumulates all submitted listwise preferences and re-trains a LightGBM model~\cite{ke_lightgbm_2017,microsoft_corporation_lightgbm_2022}. 
Each preference list is treated as one training group. 
For model training, item positions are mapped to ordinal relevance labels, leveraging the equidistant modeling assumption, so adjacent positions contribute as equally spaced preference steps. 
This simplifies the training signal, but it does not represent unequal preference gaps between neighboring items.

The retrained LightGBM ranker scores all items in the dataset; the resulting scores define the current personalized item ranking. 
To support users in inspecting and validating the ranking, Ranking Companion also computes a global ranking explanation using SHAP importance values~\cite{lundberg2017unified_SHAP}.  
The Ranking Inspection View visualizes this ranking explanation as a horizontal bar chart.


\section{Ranking Companion Interface} 
\label{sec:interface}
Ranking Companion implements the workflow through three coordinated views (Figure~\ref{fig:teaser}), each corresponding to one of the three core tasks abstracted in Section~\ref{sec:approach:workflowTasks}. 
The \textit{Item Selection Support View} (left) enables candidate selection via multiple methods (T1).
The \textit{Preference Externalization View} (center) captures listwise preferences through drag-and-drop ordering (T2).
The \textit{Ranking Inspection View} (right) displays the current ranking with lightweight explanation cues for validation and refinement (T3).


\subsection{Item Selection Support (T1)} 
\label{sec:item-selection}
This view implements five of the six item-selection methods introduced in Section~\ref{sec:approach:itemSelection}; ISM5 is integrated into the Ranking Inspection View (Section~\ref{sec:interface:result}). 
Across all methods, selecting an item adds it to the Preference Externalization View for listwise feedback submission.

\textbf{ISM0: Baseline Search}
Users retrieve known items by querying for their name and add them using (\texttt{+}). 
\texttt{Baseline Search} serves as a human-centered anchor, allowing users to seed early preferences from familiar items.

\textbf{ISM1: Similarity}
Ranking Companion also uses a row-based view for item selection; its design elements A--C are shown in Figure~\ref{fig:itemSelectionViewTypes} (left).
Users add an item with \texttt{+} (A), skip it with \texttt{x} (B) to immediately replace it with a new suggestion, and inspect top attributes via the info action (C). 
\texttt{Similarity} Suggests items similar to top-ranked items, supporting refinement around high-preference regions.

\textbf{ISM2: Trend}
Using the same row-based design, the \texttt{Trend}-based method offers items with high current listening activity relative to total play count; a dataset-intrinsic signal that introduces an external perspective independent of the user's labeled history or model state. 

\textbf{ISM3: Uncertainty}
Also using the row-based view design, \texttt{Uncertainty} prioritizes items whose attribute profiles are underrepresented in the user's labeled history, targeting regions of the attribute space where the ranking model is least informed. 

\textbf{ISM4: UMAP Projection Map}
The view (Figure~\ref{fig:itemSelectionViewTypes}, on the right) projects the full item set into 2D using UMAP~\cite{mcinnes_umap_2018}, where each point represents one item (K). 
Users pan and zoom, hover to reveal item names, and click points for item selection. 
The map supports overlap resolution in dense regions; point color encodes the last selection source, and square labels indicate top-ranked items (L). 
This method supports spatial overview and item discovery~\cite{tvcgLabeling2018}.

\textbf{ISM5: Ranking Panel}
The Ranking Inspection View on the right (Figure~\ref{fig:teaser}) presents the item ranking as a scrollable list that supports validation and correction through item selection.
Figure~\ref{fig:itemSelectionViewTypes} (center) introduces its design:
Each row enables adding items with \texttt{+} (D) and inspecting top attributes (E). 
The \texttt{Listen} action opens the item page on Last.fm (I), helping users familiarize themselves with items before adding.

\subsection{Listwise Preference Externalization (T2)}
\label{sec:interface:preferenceExternalization}
The \textit{Preference Externalization View} at the center top of Ranking Companion supports users in externalizing listwise preference feedback (Figure~\ref{fig:teaser}).
Users order a small set of previously selected items via drag-and-drop interaction.
The \emph{Submit} button below triggers model retraining.
To help users monitor their selection diversity and avoid filter-bubble effects inherent in single-source selection, Ranking Companion provides a lightweight provenance indicator that summarizes item counts per selection method as a color-coded horizontal bar chart, using the six colors consistently assigned to ISM0--ISM5 throughout the interface.


\subsection{Inspect and Validate the Ranking (T3)}
\label{sec:interface:result}

The \textit{Ranking Inspection View} supports users in inspecting and validating the current ranking.
Each row shows the rank, item name, ranking score (F), uncertainty cue (G), and a colored square indicating the last item-selection method used for the item (H), as shown in Figure~\ref{fig:itemSelectionViewTypes} (center).
On hover, users can inspect the item's top attributes on demand (E).

The Ranking Inspection View also integrates the \textit{Ranking Panel} (ISM5), enabling users to add ranked items to the preference list via the \texttt{+} action (D), inspect top attributes on demand (E), and open the item page on Last.fm (I).

The view also provides the ranking explanation.
To give users an overview of which attributes broadly influence the current ranking model, we compute SHAP importance values~\cite{lundberg2017unified_SHAP}. 
Following standard practice~\cite{lundberg2017unified_SHAP}, we derive global feature importance by aggregating attribute effects using the mean absolute SHAP value across all instances, 
and normalizing their importance scores to sum to 100\%. 
Ranking Companion visualizes this ranking explanation as a horizontal bar chart of the 10 most influential attributes. 
\looseness =-1
This gives users an interpretable signal about which attributes drive the ranking most heavily, without requiring them to interpret individual item scores.
This design follows guidance on explainable ranking, VA for XAI, and the integration of explanations into interactive workflows~\cite{hazwani_design_2022,alicioglu_survey_2022}.
We acknowledge that in a personalized, heterogeneous item space, global aggregation is a simplification: it does not capture how attribute importance may vary across different preference regions.
More localized, per-item ranking explanations remain an avenue for future work.
\section{Usage Scenarios}
\label{sec:usage-scenarios}

We illustrate Ranking Companion through two usage scenarios that follow the three-task workflow (T1--T3, Section~\ref{sec:approach:workflowTasks}) over the coordinated views in Figure~\ref{fig:teaser} and the item-selection components in Figure~\ref{fig:itemSelectionViewTypes}. 
Supplementary material with full walkthroughs and a video guide, is available \href{https://osf.io/x54t3/overview?view_only=ad39bb4b8e0446aeb4c960b4e1de581d}{online}.

\paragraph{Scenario 1 -- Refinement around known anchors}
Alex, a casual listener with a preference for electronic music, wants to build a personalized ranking for a curated playlist.
In the first iteration, Alex uses \textit{ISM0 Baseline Search} to retrieve known artists \textit{David Guetta, Diplo, and Daft Punk} and arranges them according to preference(T2).
Inspecting the updated ranking, Alex notices plausible, but unfamiliar names in the top 10 (T3) and switches to \textit{ISM1 Similarity}, which surfaces related candidates (T1) such as \textit{Martin Garrix, Alan Walker, and Edward Maya} (Figure~\ref{supp-fig:alex-02}). After enriching the initial ranking with relevant items, Alex reorders and submits this preference list (T2), and inspects the updated ranking (Figure~\ref{supp-fig:alex-03}) (T3). 

After two iterations, the provenance indicator (Figure~\ref{supp-fig:alex-04}, bottom center) shows that Alex's feedback has been drawn almost entirely from \textit{ISM0 Baseline Search} and \textit{ISM1 Similarity}. Alex turns to \textit{ISM5 Ranking Panel}, which allows items to be added directly from the current ranking (T1) to select promising but underrepresented recommendations such as \textit{Oklou, Air}, and \textit{M83} and adds them to the preference list (T1) for targeted feedback and correction (Figure~\ref{supp-fig:alex-05}) (T2).
The global SHAP ranking explanation (Figure~\ref{supp-fig:alex-07}, bottom right) helps Alex interpret how the model is using the accumulated feedback (T3). 
At this stage, the explanation in Figure~\ref{supp-fig:alex-shap-similarity} shows that the ranking is strongly driven by the attribute \textit{dance}, together with related attributes such as \textit{french}, \textit{house}, and \textit{electronica}. 
This is consistent with Alex's electronic-music intent, but it also suggests that the model is still concentrated in a narrow attribute region. 
Alex, therefore, continues using the \textit{ISM5 Ranking Panel} to add additional relevant artists from the current ranking (T1). 
After retraining, the explanation shifts toward a broader electronic profile, with \textit{electronic}, \textit{electronica}, \textit{french}, \textit{dance}, and \textit{electropop} all contributing to the learned ordering (T3) (Figures~\ref{supp-fig:alex-08} and~\ref{supp-fig:alex-shap-final}).
After three iterations, Alex has built a personalized ranking anchored around familiar electronic artists while incorporating targeted corrections from the model's current output. 
Figure~\ref{supp-fig:alex-provenance-final} shows that Alex’s final ranking was built primarily through refinement-oriented methods, while the global SHAP ranking explanation offered transparency on the attributes driving the learned ordering.

\paragraph{Scenario 2 -- Discovery across unfamiliar regions}
Sam, a musically knowledgeable user, wants to discover relevant artists beyond familiar listening habits. 
Unlike Alex, Sam does not begin by searching for known favorites, but 
construct a ranking from the ground up using complementary item-selection.
Sam first uses \textit{ISM2 Trend}, to seed the ranking (Figure~\ref{supp-fig:sam-01} with currently salient artists such as \textit{Player, Frankie Valli, and Olivia Newton-John} (T1).
Sam then turns to the \textit{ISM4 UMAP Projection Map} (Figure~\ref{supp-fig:sam-02}) to spatially browse the full item set (T1). 
By zooming into clusters distant from the initial trend-selected artists, Sam encounters candidates such as \textit{Sweet Trip, A Boogie wit da Hoodie, and Jimin}. The \texttt{Listen} action and on-demand attribute inspection help Sam judge unfamiliar artists before adding them to the preference list. 
Sam notices that \textit{Toby Fox} appears unexpectedly near the top of the ranking (Figures~\ref{supp-fig:sam-rank-trend}). 
Using hover-based attribute inspection (Figure~\ref{supp-fig:sam-03}), Sam sees that \textit{Toby Fox} is associated with \textit{soundtrack}, \textit{video game music}, and \textit{chiptune}. 
Sam therefore adds \textit{Toby Fox} from the \textit{ISM5 Ranking Panel} to reinforce this signal (T1). 
\looseness = -1
Sam then continues using the ranking panel for targeted correction, adding artists such as \textit{Tyler, The Creator, Childish Gambino, MF DOOM, Big Sean, and Earl Sweatshirt} as the ranking begins to shift toward an \textit{alternative hip-hop} and \textit{soundtrack} oriented preference region (Figure~\ref{supp-fig:sam-rank-soundtrack}). 

The provenance indicator now shows that Sam's feedback has been drawn from multiple sources, including \textit{ISM2}, \textit{ISM4}, and \textit{ISM5} (T2). 
To cover attribute regions that still remain underrepresented in the feedback history, Sam switches to \textit{ISM3 Uncertainty} (T1). 
Sam adds several artists, including \textit{The Hellp}, \textit{Jorge Ben Jor}, and \textit{Black Country, New Road} (T1), and submits a new feedback round to diversify the training signal further (T2) (Figure~\ref{supp-fig:sam-08}).

In the final Ranking Inspection View, Sam's ranking reflects a broader discovery-oriented trajectory (Figure~\ref{supp-fig:sam-01}. 
The global SHAP ranking explanation shows (Figure~\ref{supp-fig:supp-sam-shap}) that \textit{alternative hip-hop} is the strongest signal, but that \textit{underground hip-hop}, \textit{soundtrack}, \textit{rap}, \textit{electronic}, \textit{soul}, and \textit{female vocalists} also contribute to the learned ordering. 
\looseness = -1
The final ranking reflects a genuine expansion of Sam's listening horizon rather than a simple reinforcement of already-known preferences.



\section{User Study}
\label{evaluation:studyDesign}
\subsection{Methodology}
We conducted a formative user study to explore how six item-selection methods (ISM0--ISM5) differ in terms of perceived ranking quality, and (b) which methods users prefer and why. 
To answer these questions, we focused on the item selection stage. 
Our study followed a within-subject approach. 
Each participant selected items using the six methods (Section \ref{sec:approach:itemSelection}) sequentially, completing three iterations per method.  We used the $1{,}000$-artist subset of the dataset from Section~\ref{sec:approach:data}.
To ensure consistency, we asked participants to select at least three artists for each item-selection method. 
To counterbalance presentation-order effects, we used a Latin square design for presenting methods ISM1--ISM5 (Section \ref{sec:approach:itemSelection}). 
Each session began with the Baseline Search before participants proceeded through the counterbalanced order of ISM1–ISM5, to ensure a consistent starting point. 
Each session lasted about 45 minutes.

\textbf{Participants:} We recruited $N{=}10$ participants via internal university channels (ages $20$--$50$ years; 4 female, 6 male) with varying cultural backgrounds and nationalities. 
Prior artist knowledge was not a prerequisite for the study; however, we assessed their level of prior music knowledge. 
Six participants reported that they know artists across several genres.
Three participants mainly listened to a few favorite artists. 
One participant actively follows many artists across diverse genres.

\textbf{Procedure:} The study was carried out in three phases. 
First, participants gave informed consent and received a structured introduction to the tool.
They then completed a pre-study questionnaire to collect demographic data and assess prior experience. 
After the first phase, participants completed the main task. 
Following each method, participants rated six questions adapted from the ResQue questionnaire~\cite{pu_user-centric_2011} (7-point Likert scale). 
The questions cover \textit{accuracy}, \textit{diversity}, \textit{novelty}, \textit{transparency}, \textit{control}, and \textit{satisfaction}. 
Finally, we conducted a post-questionnaire using the NASA-TLX\cite{hart1988_NASA-TLX} short form, followed by two open-ended debriefing questions for further qualitative context regarding their experience. 

\subsection{Results}
\label{evaluation:results}

\begin{figure}[t]
    \centering
    \includegraphics[width=\linewidth]{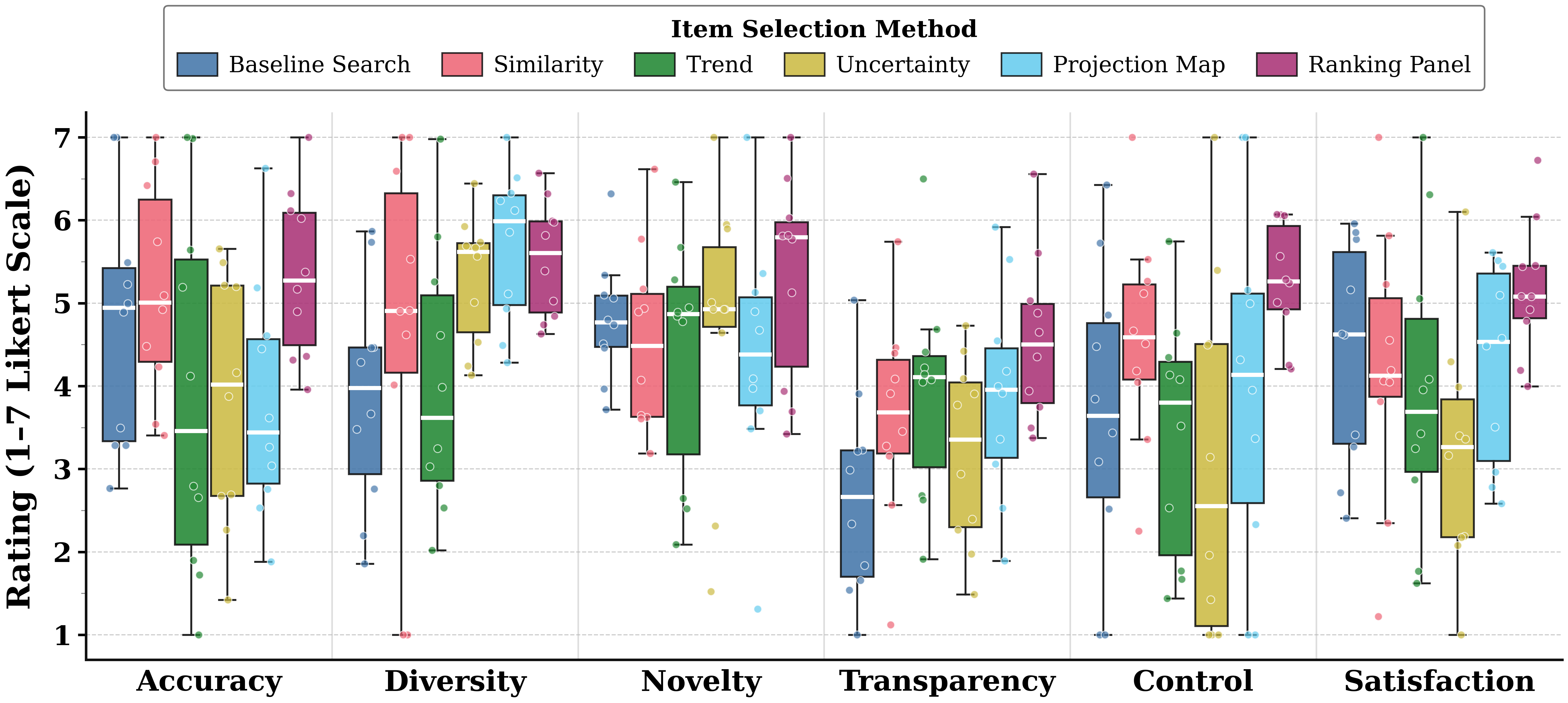} 
    \vspace{-2mm}
    \caption{Box plots of user ratings across six perceived ranking-quality dimensions show method-specific trade-offs. Ranking Panel (ISM5) selection received the highest ratings for control and satisfaction, whereas Uncertainty (ISM3) selection supported diversity but received lower transparency ratings.
    }
    \label{fig:evaluation_ratings}
    \vspace{-6mm}
\end{figure}

All participants completed the tasks in the given time frame. 
As shown in Figure~\ref{fig:evaluation_ratings}, the Baseline Search \textbf{ISM0} achieved overall satisfaction, but received lower ratings for perceived transparency and user control. 
Moreover, participants rated \textbf{ISM5}-Ranking panel selection as the most satisfactory method. 
This approach was perceived as both novel and transparent, and provided users with high perceived agency and control. 
Interestingly, participant ratings on \textbf{ISM3} Uncertainty-based ranking were mixed, with higher ratings for perceived diversity and novelty of artists, and lower ratings in terms of perceived transparency, control, and satisfaction. 

Open-ended responses provided additional context for the usage patterns reflected in Figure~\ref{fig:evaluation_ratings}. 
Participants described two strategies when selecting items: (1) \textit{refinement}, in which they first retrieved known artists to establish a baseline before using the selection methods to refine their taste profile; and (2) \textit{discovery}, in which they immediately explored unfamiliar artists to expand and shape their taste profile from the outset. 

When focusing on (1) refinement, participants found the combination of Baseline Search \textbf{ISM0} and Similarity-based  \textbf{ISM1} particularly effective for recalling familiar artists (P1, P4, P10). 
This workflow also aligned with existing streaming services, providing a sense of familiarity (P4). 
Nevertheless, participants noted limited ranking stability, with substantial shifts between iterations (P8, P9). 

When focusing on (2) discovery, participants found Trend-based ranking \textbf{ISM2} less useful, as it focused on popular items and missed niche artists (P1, P3, P4, P10). 
Instead, they found \textbf{ISM3} Uncertainty and \textbf{ISM4} UMAP Projection Map helpful. 
While Uncertainty helped suggest unfamiliar items (P3, P7), Projection fostered curiosity (P1, P4, P10). 
However, several participants noted unpredictability in the Uncertainty suggestions, decreasing perceived transparency (P2, P9, P10). 
Moreover, participants struggled with navigation and the spatial meaning of the projection (P2, P3, P6) and would have preferred a larger exploration space (P1, P5, P9).

Lastly, participants characterized \textbf{ISM5} Ranking Panel as intuitive and transparent, noting that starting from known items and allowing direct manipulation through re-ordering felt natural (P4, P10). 
Perceived limitations included a "filter-bubble" effect, where the genre narrowed focus after several iterations (P1), and increased cognitive effort, due to the multi-step nature of the feedback loop (P6).
\section{Discussion}



%




Our formative study suggests that iterative ranking creation with Ranking Companion is feasible and low-friction, but no single item-selection method received the highest ratings across all ranking-quality dimensions (Figure~\ref{fig:evaluation_ratings}). 
Instead, participants combined multiple methods as their goals shifted between preference refinement and discovery.  

\textbf{Item Selection Trade-offs:} Ranking Panel (ISM5) selection received the highest ratings for perceived accuracy, control, and transparency: by exposing the current ranking, it enabled targeted corrections and made the effects of preference feedback easier to interpret. 
However, it risks reinforcing current model preferences through filter-bubble effects. 
UMAP projection map browsing (ISM4) supported discovery without requiring name-based search and increased perceived novelty and diversity via cluster-oriented exploration. 
However, participants noted higher navigation effort and uncertainty about the spatial semantics, suggesting the need for clearer guidance and navigation support. 
AL-based item-selection methods showed mixed value: Uncertainty (ISM3) scored highly on novelty and diversity, but appeared less transparent to users, pointing to the need to improve their mental model for why items were suggested via localized, lightweight rationales. 
Trend (ISM2) suggestions primarily included popular items and were perceived as less useful for discovery, especially when participants sought niche alternatives.
Across all methods, ranking stability was a global concern: when rank shifts were large or unexplained, participants lost confidence and were discouraged from further exploration or refinement. 
Stabilizing the ranking across iterations and making rank changes interpretable may therefore be important for sustained engagement.

\textbf{Takeaways:} We take away that VA systems for ranking creation should support switching between item-selection methods as users’ shifting strategies, while making interactions accountable through clear provenance of labeled items. 
Suggested items should include local explanations about suggestion rationales, particularly for uncertainty-driven suggestions to strengthen transparency, and the interface should display ranking stability across iterations to build confidence. 
Finally, designs should mitigate filter-bubble effects when selecting from the learned ranking (ISM5) and improve navigation and interpretability in projection-map exploration (ISM4).


\textbf{Limitations:} We acknowledge several limitations with respect to our design choices and user study that point to future directions. 
First, we chose listwise, relative, equidistant ordering because it keeps preference input simple and reduces translation effort from users’ mental models to a learn-to-rank training signal \cite{xia_listwise_2008,carterette_here_2008}. 
However, this choice also constrains expressiveness, as it cannot encode preference strength or unequal distances between items; future work could explore other combinations of the design space, such as non-equidistant or absolute judgments~\cite{kuhlman_preference-driven_2018}.
Second, we designed Ranking Companion for broad audiences; a natural extension of this work is to account for user expertise and usage contexts. 
Third, our evaluation provides formative evidence, given its limitation to ten users and one domain. 
Future work could measure objective ranking accuracy or convergence through larger, controlled user studies. 
Fourth, Last.fm tag metadata includes noise and gaps in item coverage, which can affect both ranking quality and explanations.
Finally, our Uncertainty method uses an attribute-coverage heuristic as an interpretable proxy for model uncertainty instead of estimating calibrated probabilistic uncertainty or optimizing a listwise active-learning objective, and may be replaced by alternative implementations~\cite{fu_survey_2013}.

\section{Conclusion}
We presented Ranking Companion, a Visual Analytics approach for item-based ranking creation.
Its main contribution lies in its support for multiple complementary item-selection methods, with listwise preference externalization and lightweight ranking explanations. 
Our work demonstrates that \emph{item-based} preference externalization can lower the entry barrier to personalized ranking creation compared to attribute-weighting approaches, while multiple complementary selection methods can help reduce the interaction cost of producing relevant training data for learning-to-rank models. 
Our formative user study indicates that item-selection methods shape perceived ranking quality in different ways, revealing exploratory trade-offs across accuracy, diversity, novelty, transparency, control, and satisfaction.
These observations suggest that diverse item-selection strategies can help mitigate the narrowing effects of relying on a single model-driven suggestion source, supporting more transparent and controllable personalized ranking workflows.

\section{Acknowledgements}
This work was supported by the Swiss National Science Foundation (SNSF), grant no. 10003068 on Personalized Visual Analytics: Human Preference Elicitation for Ranking-based Multi-Criteria Decision-Support.

\bibliographystyle{IEEEtran}
\bibliography{thebibliography}

\end{document}